\documentclass[10pt,a4paper]{article}

\usepackage{amsmath}
\usepackage{amsfonts}
\usepackage[utf8]{inputenc}
\usepackage{subfigure}
\usepackage{graphicx}  
\usepackage{authblk}

\title{Carbon nanotubes as target for directional detection of light WIMP}
\author[1,2]{\small V.C.~Antochi}
\author[3]{\small E.~Baracchini}
\author[1,2]{\small G.~Cavoto}
\author[2]{\small E.~Di Marco}
\author[3]{\small G.~Mazzitelli}
\author[2]{\small D.~Pinci}
\author[1,2]{\small A.D.~Polosa}
\author[2]{\small F.~Renga}
\author[2]{\small C.~Voena}
\date{}
\affil[1]{\footnotesize Dipartimento di Fisica, Universit\`a di Roma - Roma, Italy}
\affil[2]{\footnotesize INFN, Sezione di Roma - Roma, Italy}
\affil[3]{\footnotesize INFN, Laboratori Nazionali di Frascati - Frascati, Italy}

\begin{document}

\maketitle

\begin{abstract}
In this paper I will briefly introduce the idea of using Carbon Nanotubes (CNT) as target for the detection of low mass WIMPs with the additional information of directionality. I will also present the experimental efforts of developing a Time Projection Chamber with a CNT target inside and the results of a test beam at the Beam Test Facility of INFN-LNF.
\end{abstract}

\section{The idea}
As the Solar System moves through the Galaxy, if we assume a dark matter halo, as astrophysical evidence suggests, made of cold weakly interacting massive particles (WIMPs), we expect a wind of these particles coming from the direction towards the Sun is moving, the Cygnus constellation. The Earth's revolution around the Sun would also be expecting to have a consistent effect of modulating a possible signal of dark matter, as seen by Dama/LIBRA experiment at INFN Gran Sasso, Italy~\cite{ref:DAMA}. So in order to assure the nature of possible interactions due to WIMPs, a directional information of these events is needed.\\
In order to solve the problem of directional detection of dark matter we explore the possibility of using a large array of alined Carbon Nanotubes (CNTs) open at one end as target inside a Time Projection Chamber (TPC). The concept is that a WIMP interacts with a carbon nucleus inside a CNT elastically and then the recoiling nucleus, if the recoil direction is close to the CNT axis, has a good chance to be channeled towards the open cap and inside the gaseous TPC~\cite{ref:CNT1}\cite{ref:CNT2}\cite{ref:CNT3}.\\
The ideal detector would be a TPC with a CNT wall inside with the axes of the nanotubes directed towards Cygnus at all times and a drift field perpendicular to the CNT axes.
The objective for this kind of detector is to be able to observe recoil events down to a threshold of 1 KeV. This way assuming no signal events and rejection of all background, in 3 years of exposure time and a 10kg CNT target the detector would reach a cross section sensitivity of $10^{-44}$cm$^2$ for WIMPs of mass 1-2 GeV.

\section{The experimental set-up}
A prototype was developed in order to study the CNT characteristics and response. This prototype consisted in a 3D-printed field cage with a 5 cm drift gap, with silver rings spaced 1 cm apart in order to mantain a uniform drift field. Triple thin GEMs~\cite{ref:GEM}, 3cm$\times$3cm wide, were used as amplification and a TimePix chip~\cite{ref:TIMEPIX} was used as read-out to collect the charge from the GEMs.\\
The CNT target used was a 2cm$\times$2cm silicon substrate with 200$\mu$m thick multi wall carbon nanotube forest on one side, grown perpendicularly. The target was placed vertically in a minimal 3D-printed plastic support inside the drift gap.\\
The first tests with the CNT target were performed at the Beam Test Facility (BTF)~\cite{ref:BTF} at LNF. The TPC was placed inside an inox steel vacuum vessel whith a thin Berillium window for the beam, the vessel was set on the micrometric table of the BTF. This way beam height could be easilly controlled remotely. The vessel was then filled with different gas mixtures for different performance tests, but the main results were achieved with the Ar:CO$_2$:CF$_4$ (40:55:5) and pure SF$_6$ at low pressures.\\
The BTF beam used was 450 MeV electron beam with the possibility to modify its intensity and its dimensions, and to monitor it directly with a MediPix chip. 

\section{First results}
\begin{figure}[!b]
	\centering
	\subfigure[]{\includegraphics[width=.49\textwidth]{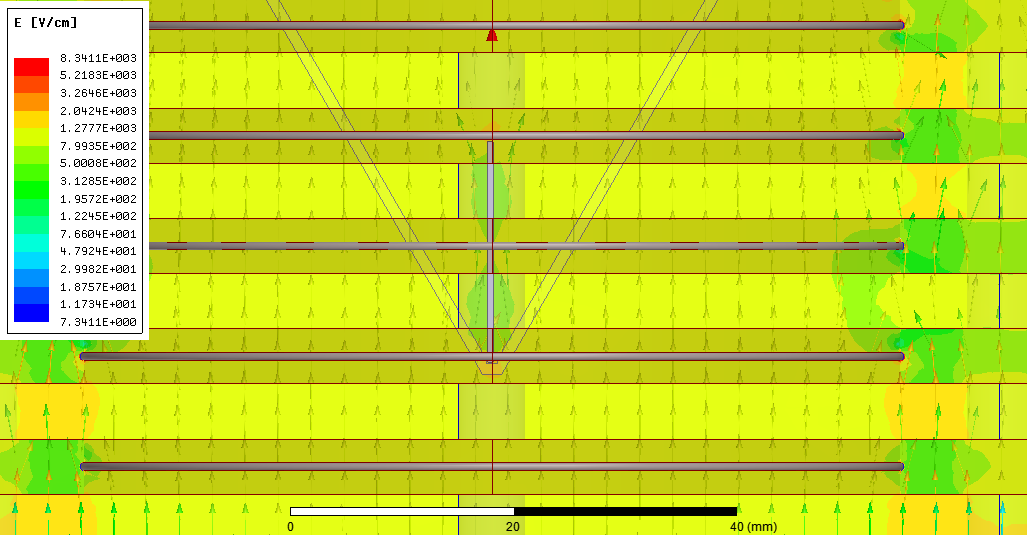}}
	\subfigure[]{\includegraphics[width=.49\textwidth]{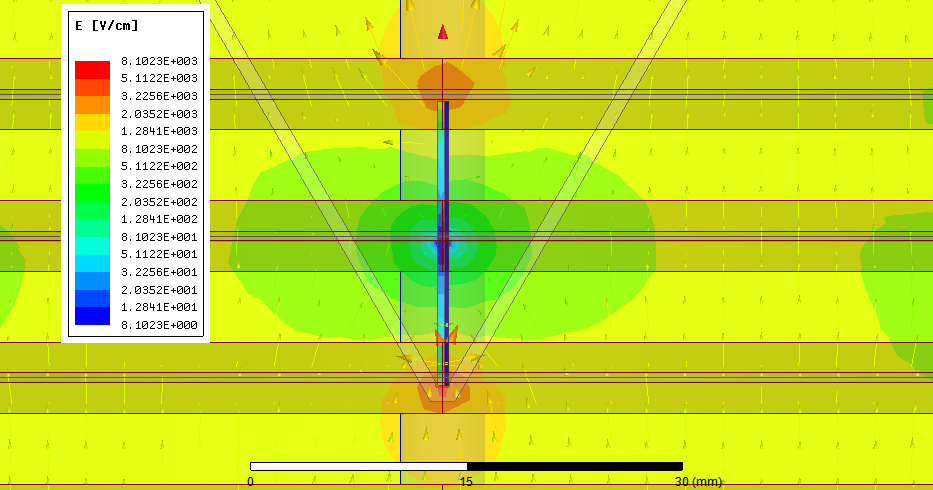}}   
	\caption{\label{fig:SIM} Simulation in ANSYS Maxwell of the TPC with the target of semiconductive Si substrate only (a), and Si + a sheet of conductive graphite to resemble the CNTs (b). The drift field (intensity indicated by colour gradient and direction by arrows) in (a) shows very small distortion effects while in (b) it is very distorted by the conductive graphite.}
\end{figure}
\subsection{Drift field simulation}
Before the experimental tests, we performed simulations with the ANSYS Maxwell software of the entire geometry of the TPC in order to verify the uniformity of the drift field and study possible distortions introduced by the CNT target. The CNT target, allegedly semiconductive material grown onto a silicon (semiconductive) base, was simulated first as a semiconductor, then as a conductive graphite sheet onto a semiconductive silicon base.\\
In the first case the simulation showed very little distortion of the drift field close to the edges of the semiconductive target, but in the second case the graphite distorted heavilly the drift field (see Fig.~\ref{fig:SIM}).\\

\subsection{Experimental results}
\begin{figure}[!t]
	\centering
	\includegraphics[width=.6\textwidth]{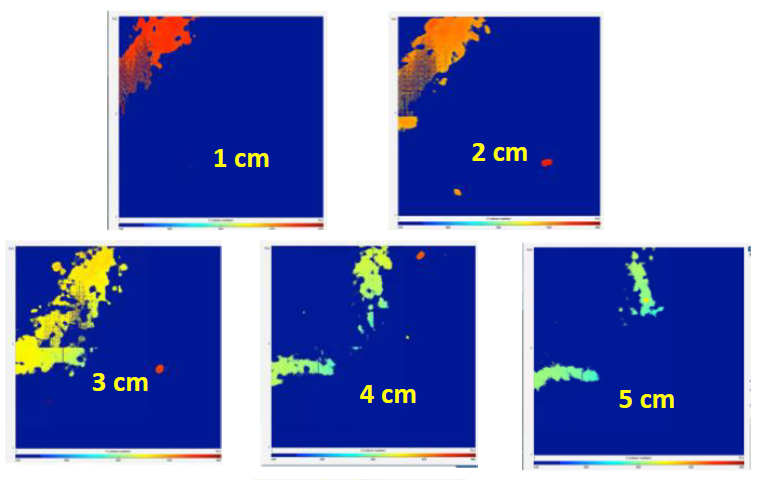}     
	\caption{\label{fig:dist1}Images of the projection of the beam passing through the field cage at different heights from the base. The beam is straight at 1cm and 2cm from the base of the drift gap, but when getting closer to the CNT target it shows distortions of the drift field (3, 4 and 5cm). The different colours, red to blue, are proportional to the height of the beam and measure the time difference between the time of arrival and the LINAC's t$_0$.}
\end{figure}
At the BTF the configuration with the vertical CNT target was tested. The electron beam was sent at various heights through the drift gap in order to study the response of the CNTs when the beam passed close to them. The TimePix chip recorded the projection of the ionization produced by the beam in the TPC. Below the target the beam projection was recorded straight wihtout any distortion effect from the CNTs, but going higher and closer to the target, strong distortion effects were observed as seen in Fig.~\ref{fig:dist1}, comparable to those seen in the simulation with the conductive graphite sheet.
The experimental behaveour of the CNT target, in a drift field of $\sim$0.8kV/cm in both Ar:CO$_2$:CF$_4$ and SF$_6$, resembled more a conductor than a semiconductor.\\
New configurations are being studied in order to minimise the field distortion and new tests will be performed at the BTF in July 2017.\\




%


\end{document}